\documentclass[pdftex,twocolumn,epjc3]{svjour3}          

\RequirePackage[T1]{fontenc}

\smartqed  

\RequirePackage{graphicx}
\RequirePackage{mathptmx}      
\RequirePackage{flushend}
\RequirePackage[numbers,sort&compress]{natbib}
\RequirePackage[colorlinks,citecolor=blue,urlcolor=blue,linkcolor=blue]{hyperref}

\journalname{Eur. Phys. J. C}

\begin{document}

\title{Gravitational fields and harmonic gauge in brane model of Universe.\thanksref{t1}}


\author{Sergey N. Andrianov\thanksref{e1,addr1}
        \and
        Rinat A. Daishev \thanksref{e2,addr2}
                \and
        Sergey M. Kozyrev  \thanksref{e3,addr3}
        \and
        Boris P. Pavlov  \thanksref{e4,addr4} 
}

\thankstext[$\star$]{t1}{Thanks to the title}
\thankstext{e1}{e-mail: adrianovsn@mail.ru}
\thankstext{e2}{e-mail: rinat.daishev@mail.ru}
\thankstext{e3}{e-mail: kozyrev@dulkyn.ru}
\thankstext{e4}{e-mail: pavlov.gpb@gmail.com}

\institute{Tatarstan Academy of Sciences, Institute of Applied
Research 420111, 20, Bauman str. Kazan Russia\label{addr1}
          \and
          Kazan Federal University, Institute of Physics,
420008, 18 Kremlyovskaya str. Kazan Russia \label{addr2}
          \and
          Scientific Center for Gravity Wave Studies "Dulkyn",
420036, 12-5, Lyadova str. Kazan, Russia \label{addr3}
          \and
          Kazan National Research Technical University named after A. N. Tupolev - KAI,
420111 10, Karl Marx Str. Kazan, Russia \label{addr4} }

\date{Received: date / Accepted: date}

\maketitle

\begin{abstract}
We suppose that our Universe is closed manifold in real embedding
higher dimensional space. This model well describes expanding
character of Universe where each point becomes more far from any
other point with time. We have derived Klein-Gordon equation using
the symmetry of Universe expansion. Comparing it with squared
Dirac-Fock-Ivanenko equation we have derived expression for
gravita- tionally induced chiral current. Starting from the
expression for gravitationally induced chiral current obtained on
the basis of brane symmetry we have shown that being placed into
the Einstein's gravitational field, chiral current yields inertial
gravitational field. This inertial gravitational field transforms
into the vortex gravitational field and then back to the inertial
gravitational field thus ensuring the propagation of such kind of
gravitational excitation as constitutional part of gravitational
wave. In the case of weak gravitation, these symmetry
considerations and our consequent equations lead directly to
harmonic gauge conditions that are necessary for supplementing
Einstein equations.
\end{abstract}

\section{Introduction}

Brane can be regarded as multidimensional spatial manifold
supporting the existence of particle in Universe. Zero dimen-
sional brane is a point particle, one dimensional brane is a
string. Brane with p dimensions is called p-brane. Open string
describing the particle living on brane can be connected by its
ends to brane and the brane, in this case, is called D-brane.
Initially, dimensions of brane were considered as small scale ones
but, in 1983, V.A. Rubakov and M.E. Shapo- shnikov had introduced
the model of brane in large additional dimensions \cite{Rubakov}.
Branes like strings \cite{Dai}, \cite{Horava}, \cite{Polchinski},
\cite{Horava2} can be closed manifolds. Already Alexander
Friedmann considered the whole Universe in particular variant of
his models as closed surface but this surface was regarded as
hypothetical one \cite{Friedmann}. Eventually, Merab
Gogberashvilly had proposed the closed brane model of Universe as
real spherical expanding shell \cite{Gogberashvili}. He had
introduced the model where our Universe is considered as an
expanding four-dimensional bubble in five dimensions with a center
of Big Bang in the fifth dimension. This model well describes
expanding character of Universe where each point becomes more far
from any other point with time in the course of galaxies runaway.
In addition, this model describes well the isotropic character of
relict background radiation in our Universe. Besides, this model
can explain why Lorenz invariance is realized on 4d shell while it
is lost in the fifth dimension as it was known earlier in
extra-dimensional compactification theories \cite{Rizzo},
\cite{Garcia}. As an alternative to compactification theories,
Randal and Sun- drum had proposed semi-phenomenological brane
model \cite{Randall} where it was shown that four-dimensional
Newton and Ein- stein gravity live in five dimensional space.
Cosmological extensions of this model were given soon in papers
\cite{Mohapatra}, \cite{Cline}, \cite{Mukohyama}, \cite{Flanagan},
\cite{Cline2}. We have derived Klein-Gordon equation using the
symmetry of Universe expansion \cite{Andrianov},
\cite{Andrianov2}. Com- paring it with squared Dirac-Fock-Ivanenko
equation, we have derived expression for gravitationally induced
chiral current \cite{Andrianov3}. Here, we show that being placed,
in its turn, into the Einstein's gravitational field, chiral
current yields inertial gravitational field. This inertial
gravitational field transforms into the vortex gravitational field
and then back to the inertial gravitational field thus ensuring
the propagation of these fields along with gravitational wave. In
addition, we analyze our equations in the limit of weak
gravitation and show that harmonic gauge often using with
Einstein's equations is a consequence of brane rotational
symmetry.

\section{Gravitational fields equations}
\label{sec:1} Symmetry of Universe in brane model regarding its
expansion yields Klein Gordon equation for a particle with mass m
and wave function   moving on brane \cite{Andrianov},
\cite{Andrianov2}:

\begin{equation}
\left\{ g^{\mu\nu} \nabla_\mu \nabla_\nu+\frac{1}{4}R
+\left(\frac{m c}{\hbar}\right)^2\right\}\Psi=0
 \label{eqKlein Gordon}
\end{equation}
where $g^{\mu\nu}$  is metric tensor, $\nabla_\mu, \nabla_\nu $
are Riemann connection covariant derivatives and $R$ is scalar
curvature, Greek indices run from 0 to 3. Comparison of this
Klein-Gordon equation with squared Dirac-Fock-Ivanenko equation
\begin{eqnarray}\label{eqDirac-Fock-Ivanenko}
i \gamma^\mu(\nabla_\mu +\Gamma_\mu) \psi = m \psi
\end{eqnarray}
where $ \gamma^\mu$  is Dirac matrix and $\Gamma_\mu$  is spin
connection, yields the following relation:
\begin{equation}\label{eq3}
 \gamma^\mu \gamma^\nu D_\mu \Gamma_\nu =- \frac{1}{4}R
\end{equation}
where $D_\mu = \nabla_\mu+ \Gamma_\mu$ is covariant derivative
generalized on spin connection. In its turn, relation (\ref{eq3})
leads to the following expression for determination of
gravitationally induced chiral current  $j^{\alpha 5}$:
\begin{equation}\label{eq4}
j^{\alpha 5} \varepsilon ^{\mu\nu\tau\xi} R_{\mu\nu\tau\xi} =4 i
\gamma ^ \alpha D_\mu \Gamma^\mu
\end{equation}
where $\varepsilon ^{\mu\nu\tau\xi}$  is Levi-Civita tensor,
$R_{\mu\nu\tau\xi}$ is Riemann curvature tensor, $\gamma ^ \alpha$
is Dirac matrix, $\Gamma^\mu =
g^{\alpha\beta}\Gamma^{\mu}_{\alpha\beta}$ is spin connection,
$g^{\alpha\beta}$ is metric tensor, $\Gamma^{\mu}_{\alpha\beta}$
is Christoffel symbol. By multi- plication of expression
(\ref{eq4}) both parts on Levi-Chevita tensor
\begin{equation}
\gamma^5 \frac{1}{4!}\varepsilon_{\zeta \lambda \chi
\sigma}\varepsilon^{\mu\nu\tau\xi}R_{\mu\nu\tau\xi}=
\frac{1}{4!}4i \varepsilon_{\zeta \lambda \chi \sigma}
\gamma^\alpha D_\mu \Gamma^\mu\label{eq5}
\end{equation}
we come to the expression
\begin{equation}
j^{\alpha 5} \delta^{\mu \nu \tau \xi}_{\zeta \lambda \chi \sigma}
R_{\mu\xi \nu \tau}=\frac{i}{6}\varepsilon_{\zeta \lambda \chi
\sigma}\gamma^\alpha D_\mu \Gamma^\mu , \label{eq6}
\end{equation}
and then to
\begin{equation}
j^{\alpha 5} R_{\zeta \lambda \chi \sigma}=\varepsilon_{\zeta
\lambda \chi \sigma}j^\alpha, \label{eq7}
\end{equation}
where
\begin{equation}
j^{\alpha} =\frac{i}{6}\gamma^\alpha D_\mu \Gamma^\mu, \label{eq8}
\end{equation}
or alternatively to
\begin{equation}
j^{\alpha 5} R_{\lambda\sigma}=\varepsilon_{\zeta \lambda \chi
\sigma}g^{\zeta\chi} j^\alpha , \label{eq9}
\end{equation}
by using relations $ R_{\lambda\sigma}=g^{\zeta \chi} R_{\zeta
\lambda \chi \sigma}$  and $R=g^{\lambda\sigma}R_{\lambda\sigma}$.
What is for $R_{\lambda\sigma}$ it can be found from Einstein's
equation
\begin{equation}
R_{\lambda\sigma}-\frac{1}{2}g_{\lambda\sigma}R=\frac{4\pi
G}{c^4}T_{\lambda\sigma} , \label{eqEinstein}
\end{equation}

Starting from the expression
\begin{equation}\label{eq11}
 \gamma^\mu \gamma^\nu D_\mu \Gamma_\nu =- \frac{1}{4}R
\end{equation}
we get using relation

\begin{eqnarray}
&&
\begin{tabular}{l}
$\gamma^\mu \gamma^\nu D_\mu \Gamma_\nu =\frac{1}{2}
\left\{\gamma^\mu, \gamma^\nu\right\}D_\mu \Gamma_\nu
+ $ \\
\\
$+\frac{1}{2}\left[\gamma^\mu, \gamma^\nu
\right]\left(\frac{1}{2}(D_\mu \Gamma_\nu+ D_\nu
\Gamma_\mu)+\frac{1}{2}(D_\mu \Gamma_\nu- D_\nu
\Gamma_\mu)\right).$
\end{tabular}
\label{eq12}
\end{eqnarray}
the following equation
\begin{equation}\label{eq13}
D_\mu \Gamma^\mu \frac{1}{2} \gamma_\mu \Gamma_\xi (D_\mu
\Gamma_\xi- D_\xi \Gamma_\mu)=- \frac{1}{4}R
\end{equation}

Finally, we have the following set of equations
\begin{equation}
R_{\lambda\sigma}-\frac{1}{2}g_{\lambda\sigma}R=\frac{4\pi
G}{c^4}T_{\lambda\sigma} , \label{eqEinstein2}
\end{equation}
\begin{equation}
j^{\alpha 5} R_{\lambda\sigma}=\frac{i}{6}\varepsilon_{\zeta
\lambda \chi \sigma}g^{\zeta\chi} \gamma^\alpha D_\mu \Gamma^\mu,
\label{eq15}
\end{equation}

\begin{equation}\label{eq16}
D_\mu \Gamma^\mu \frac{1}{2} \gamma_\mu \Gamma_\xi (D_\mu
\Gamma_\xi- D_\xi \Gamma_\mu)=- \frac{1}{4}R
\end{equation}

The left side of equation (\ref{eqEinstein2}) can be expressed
solely through metric tensor and thus can serve in finding it via
known energy-momentum tensor. In other words, energy dis- turbance
produces Einstein's metric gravitational field. It con- sists of
the part connected with Ricci tensor curvature $R_{\lambda\sigma}$
and the part connected with scalar curvature $R$. In its turn, the
tensor curvature part of field produces via the left side of
equation (\ref{eq15}) gravity inertial field $D_\mu \Gamma^\mu)$
in its right part. Gravity inertial field is expressed through
spin connection $\Gamma^\mu$ and is related to scalar curvature by
equation (\ref{eq16}).  This gravity inertial field turns into
gravity vortex field and vice versa ensuring propagation of this
type gravitational fields according to this equation. It can be
accomplished through creation and annihilation of virtual
particles. This propagation is absorption less if the scalar
curvature is constant. Thus, this set of equations completely
describes the propagation of gravitational fields where
gravitational wave is described by shown complex structure of
gravitational fields supplementing Einstein equations up to full
set.

\section{ Harmonic gauge.}
\label{sec:2} Let's consider equation (\ref{eq16}) in the case of
weak gravitation. It can be rewritten as
\begin{eqnarray}
&&
\begin{tabular}{l}
$\gamma^\mu \gamma^\nu g_{\nu\sigma} [g^{\alpha\beta}
g^{\xi\lambda}\frac{\partial}{\partial x_\mu} \left(\frac{\partial
g_{\lambda\alpha}}{\partial x_\beta}+\frac{\partial
g_{\lambda\beta}}{\partial x_\alpha}-\frac{\partial
g_{\alpha\beta}}{\partial x_\lambda} \right) + $ \\
\\
$+\left(\frac{\partial g^{\alpha\beta}}{\partial x_\mu}+
g^{\alpha\beta} \frac{\partial g^{\xi\lambda}}{\partial
x_\mu}\right)\left(\frac{\partial g_{\lambda\alpha}}{\partial
x_\beta}+\frac{\partial g_{\lambda\beta}}{\partial
x_\alpha}-\frac{\partial g_{\alpha\beta}}{\partial x_\lambda}
\right)+ $ \\
\\
$ +
4\Gamma^{\xi}_{\kappa\nu}\Gamma^{\kappa}]+4\Gamma_{\mu}\Gamma^{\xi}=R.$
\end{tabular}
\label{eq17}
\end{eqnarray}
Keeping in mind that
\begin{equation}
\Gamma^{\xi}_{\kappa\nu}=\frac{1}{2}g^{\xi\lambda}\left(\frac{g_{\lambda\kappa}}{\partial
x_\mu}+\frac{g_{\lambda\mu}}{\partial
x_\kappa}-\frac{g_{\kappa\mu}}{\partial x_\lambda}\right),
\label{eq18}
\end{equation}
and
\begin{equation}
\Gamma^\kappa=\frac{1}{2}g^{\alpha\beta}\Gamma^{\kappa}_{\alpha\beta},
\label{eq19}
\end{equation}
we introduce linear approximation
\begin{equation}
g_{\mu\nu}(x)=\eta_{\mu\nu}+h_{\mu\nu}. \label{eq20}
\end{equation}
where $\eta_{\mu\nu}$  is pseudo-Euclidian metrics. Assuming that
not only $h_{\mu\nu}$  is small but, also, its derivatives we have
at vacuum conditions $(R=0 )$ from (\ref{eq17}):

\begin{equation}
\gamma^\mu\gamma^\nu\frac{\partial}{\partial
x_\mu}\left(\frac{\partial h^\alpha_\nu}{\partial
x_\alpha}-\frac{1}{2}\frac{\partial h}{\partial x_\nu} \right)=0.
\label{eq21}
\end{equation}
or
\begin{equation}
\gamma^\mu\gamma^\nu\frac{\partial V_\nu}{\partial x_\mu}=0
\label{eq22}
\end{equation}
The only solution of equation (\ref{eq22}) is trivial one
\begin{equation}
V_\nu=\frac{\partial h^\alpha_\nu}{\partial
x_\alpha}-\frac{1}{2}\frac{\partial h}{\partial x_\nu} =0,
\label{eq23}
\end{equation}
that is harmonic gauge. Thus, we had shown that harmonic gauge is
a consequence of brane symmetry.

\section{ Conclusion.}
\label{sec:3}Thus, we have derived Klein-Gordon \cite{Ta-Pei},
\cite{Weise} on equation for a quantum particle starting from the
symmetry properties of the brane. This had confirmed the validity
of discussed above spherical model of Universe. Comparison of this
Klein-Gordon equation with the squared Dirac-Fock-Ivanenko equa-
tion yielded expression for chiral current that had given rise to
chiral symmetry breaking at early stages of universe evo- lution
and creation of particles baryonic mass constituting 99{\%} of
visible matter in Universe. It is shown that this chiral current
serves as a source for special new type of gravitational fields
that propagate as constitutional part  of gravitational wave
described by Einstein's equation. These type gravita- tional
fields propagate by transforming of gravitational inertial field
into vortex field and vice versa through creation and annihilation
of virtual particles. We had shown that obtained equations
describe in the case of weak gravitation harmonic gauge relations
that are usually supplement to Einstein equa- tions making them a
closed set. Thus, harmonic gauge is the consequence of brane
rotational symmetry in additional dimension space. This rotation
symmetry provides invariance of our four-dimensional space located
on four dimensional brane immersed in five dimensional universal
space. This five dimensional  space has preferable frame with the
center of coordinates in the point of Big Bang serving as a center
of above said rotation.  Recent observations of gravitational
waves \cite{Abbott} confirm validity of harmonic gauge because it
is the gauge leading to the wave like form of propagating
gravitational excitations. Therefore, they confirm also  the
existence of Universal brane as a spherically symmetric shell in a
higher dimensional space. Some recent theoretical papers consider
a number of such symmetry consequences \cite{Belayev},
\cite{Bobev}, \cite{Troisi}, \cite{Banerjee}. In particular, the
upgrade of this model was presented where the effect of mass was
considered as a localized brane deformation imparted by the
endpoint of the string stretched in additional dimension
\cite{Banerjee}.


\begin{thebibliography}{9}


\bibitem{Rubakov} V. A. Rubakov and M. E. Shaposhnikov. \emph{Do we live inside a domain wall?} Physics Letters B, V.125, 136-138 (1983).
\bibitem{Dai} J. Dai, R.G. Leigh and J. Polchinski. \emph{New connections between string theories}. Mod. Phys. Lett. A 4,  2073-2083 (1989).
\bibitem{Horava} Horava, P. \emph{Strings on Worldsheet Orbifolds}, Nucl. Phys. B327, 461-484, (1989).
\bibitem{Polchinski} J.Polchinski, \emph{Dirichlet Branes and
Ramond-Ramond Charges}. Phys. Rev. Lett. 75, 4724 (1995).
\bibitem{Horava2} P.Horava, , E.Witten,  \emph{Heterotic and Type I String Dynamics from
       Eleven Dimensions}, Nucl. Phys. B460, 506-524 (1996).
\bibitem{Friedmann} A. Friedmann. \emph{\"{U}ber die Kr\"{u}mmung des Raumes}. Zeitschrift f\"{u}r Physik A, V.10, No.1, P.377-386 (1922).
\bibitem{Gogberashvili} M. Gogberashvili. \emph{Our World as an Expanding Shell}. Europhys. Lett., 49, 396 (2000).
\bibitem{Rizzo} Th.G. Rizzo. \emph{Lorentz Violation in Extra Dimensions.} Journal of High Energy Physics,
JHEP 09 036 (2005); hep-ph/0506056v2.
\bibitem{Garcia} J.D. Garcia-Aguilar, A.
Perez-Lorenzana \emph{Superfield formalism for 5D Lorentz
violating models.} Journal of Physics: Conference Series, V.378,
012004 (2012).
\bibitem{Randall} L. Randall, R. Sundrum. \emph{ An alternative to
compactification.}, Phys. Rev. Lett. V.83, P.4690 (1999).
\bibitem{Mohapatra} R.N. Mohapatra, A. Perez-Lorenzana, and C. A. de Sousa Pires.
\emph{Cosmology of brane-bulk models in five dimensions. } Int. J.
Mod. Phys. A, V.16, P.1431 (2001). hep-ph/0003328.
\bibitem{Cline} J. M. Cline. \emph{ Cosmological
expansion in the Randall-Sundrum warped
compactification.}Proceedings of the Third International Workshop
on Particle Physics and the Early Universe. Cosmo99, P.472-479.
hep-ph/0001285.
\bibitem{Mukohyama} S. Mukohyama, T. Shiromizu, and K. Maeda.
\emph{Global structure of exact cosmological solutions in the
brane world.} Phys. Rev. D, V.62, 024028 (2000); hep-th/9912287.
\bibitem{Flanagan} E. E. Flanagan, S. H. Tye, and I. Wasserman. \emph{Cosmological expansion in the Randall-Sundrum
brane world scenario.} Phys. Rev. D, V.62, 044039 (2000);
hep-ph/9910498.
\bibitem{Cline2} J. M. Cline, C. Grojean, and G. Servant. \emph{ Cosmological expansion in the presence of
extra dimensions.} Phys. Rev. Lett. V.83, 4245 (1999);
hep-ph/9906523.
\bibitem{Andrianov}   S.N. Andrianov, V.V. Bochkarev.\emph{Quantum equation of motion for
a particle in the field of primordial fluctuations.} Proc. SPIE,
V.7024, International Workshop on Quantum Optics 2007, 70240Z (17
April 2008).
\bibitem{Andrianov2}  S.N. Andrianov, R.A.
Daishev, S.M. Kozyrev.  \emph{Klein-Gordon equation for a particle
in brane model} Global Journal of Science Frontier Research.
Physics and Space Science, V.13, 1-5 (2013).
\bibitem{Andrianov3} S.N. Andrianov, R.A. Daishev, S.M. Kozyrev.
\emph{Gravitationally induced chiral current and mass generation.}
Hadronic Journal, V.39, 300-307 (2016).
\bibitem{Ta-Pei}  Ta-Pei Cheng and Ling-Fong Li.  \emph{Gauge Theory of Elementary Particle
Physics,}  (Oxford 1984) ISBN 978-0198519614.
\bibitem{Weise} W. Weise. \emph{Nucleon mass: From lattice QCD to the chiral limit.}
 Physical Review, V.D73 114510 (2006).
\bibitem{Abbott} B.P. Abbott et al. (LIGO Scientific Collaboration and Virgo
Collaboration): \emph{Observation of Gravitational Waves from a
Binary Black Hole Merger.} Phys. Rev. Lett. 116, 061102(1-16)
(2016).
\bibitem{Belayev}  W. Belayev.  \emph{Central Field in Rotating Spherical Space in 5D.} Frontiers in Science, V.2(6), P.159-168 (2012).
\bibitem{Bobev} N. Bobev, P. Bomans and F.F. Gautason. \emph{Spherical Branes.}
Journal of High Energy Physics, V.8 029 (2018).
\bibitem{Troisi} Antonio Troisi. \emph{Higher-order gravity in higher dimensions: geometrical origins
of four-dimensional cosmology?} Eur. Phys. J. C, V.77, 171 (2017).

\bibitem{Banerjee} S. Banerjee, U. Danielsson, G. Dibitetto, S.
Giri, and Marjorie Schillo. \emph{ Emergent de Sitter Cosmology
from Decaying Anti-de Sitter Space.}  Phys. Rev. Lett., V.121,
261301 (2018).

\end{thebibliography}
\end{document}